\documentclass[usenatbib]{mn2e}
\usepackage{amssymb,amsmath}
\usepackage{graphicx,times}
\usepackage{latexsym}
\usepackage{color}
\usepackage{array}

\def\dcr{d_{\rm hard}}
\def\dbh{d_{\rm bh}}
\def\Rcr{R_{\rm cr}}
\def\fBH{f_{\rm BH}}
\def\Sigmacr{\Sigma_{\rm cr}}

\def\kms{\,{\rm km\,s^{-1}}}
\def\muth{\mu_{\rm th}}

\title[Effects of Supermassive binary black holes on galaxy lenses]
{Effects of supermassive binary black holes on gravitational lenses}
\author[Li, Mao, Gao, Loeb \& Di Stefano]{Nan Li $^{1}$\thanks{E-mail:uranus@bao.ac.cn},
Shude Mao$^{1, 2}$, 
Liang Gao$^{1}$, 
Abraham Loeb$^3$, R. Di Stefano$^{3}$ \\ 
$^{1}$National Astronomical Observatories, Chinese Academy of
Sciences, 20A Datun Road, Beijing 100012, China \\
$^{2}$Jodrell Bank Centre for Astrophysics, University of Manchester, M13 9PL, UK \\
$^{3}$Harvard-Smithsonian Center for Astrophysics 60 Garden Street, MS-5, Cambridge, MA 02138, USA}
\begin{document}
\maketitle
\begin{abstract}
Recent observations indicate that many if not all galaxies host massive central black
holes (BHs). In this paper we explore the influence of supermassive binary black holes (SMBBHs) 
on their actions as gravitational lenses. When lenses are modelled as singular isothermal ellipsoids, 
binary black holes change the critical curves and caustics differently as a function of distance.
Each black hole can in principle create at least one additional image, which, if observed, provides
evidence of black holes. 
By studying how SMBBHs affect the cumulative distribution of magnification for images created by black holes,  
we find that the cross section for at least one such additional image to have a magnification larger than $10^{-5}$
is comparable to the cross section for producing multiple-images in singular isothermal lenses.   
Such additional images may be detectable with high-resolution and 
large dynamic range maps of multiply-imaged systems from future facilities, such as the SKA. 
The probability of detecting at least one image (two images) with magnification above 
$10^{-3}$ is $\sim 0.2 \fBH$ ($\sim 0.05 \fBH$) in a multiply-imaged lens system,  
where $\fBH$ is the fraction of galaxies housing binary black holes. 
We also study the effects of SMBBHs on the core images when galaxies have shallower central density profiles 
(modelled as non-singular isothermal ellipsoids). 
We find that the cross section of the usually faint core images is further suppressed by SMBBHs. 
Thus their presence  should also be taken into account when one constrains the core radius from the lack of central images in gravitational lenses.

\end{abstract}
\section{introduction}
Recent observations suggest that many if not all nearby
galaxies host massive central black holes. Empirical correlations
have been discovered between the mass of the supermassive black hole (SMBH) and
various galaxy properties such as the bulge mass \citep{2001ApJ...553..677L, 2003ApJ...589L..21M, 2004ApJ...604L..89H, 
2006ApJ...637...96N, 2009MNRAS.398L..41S}, 
velocity dispersion \citep{2000ApJ...539L...9F, 2000ApJ...539L..13G, 2002ApJ...574..740T, 
2003MNRAS.342..501N, 2006ApJ...641...90R, 2008MNRAS.386.2242H, 2011MNRAS.412.2211G, 2011GReGr..43.1007F}, 
luminosity \citep{1998AJ....115.2285M, 2001MNRAS.327..199M, 2002MNRAS.331..795M, 2007MNRAS.379..711G} 
and concentration \citep{2001ApJ...563L..11G, 2007ApJ...655...77G}. 
These correlations suggest that the growth of black hole is closely related to galaxy formation 
(e.g., \citealt{2000MNRAS.311..576K, 2000MNRAS.311..279M, 2002ApJ...581..886W, 2002MNRAS.335..965Y,  
2003ApJ...582..559V, 2003ApJ...593...56D, 2004ApJ...606..763H,2006ApJ...644L..21F, 2008ApJ...689..732Y, 2009ApJ...704.1135B}), although such a relation can also 
be a consequence of the central limit theorem in galaxy mergers with no significant physical meaning \citep{Peng2007}.

Gravitational lensing is an independent, mass-based method to probe SMBHs. 
The lensing effect of single central and off-centre SMBHs have been studied previously
\citep{2001MNRAS.323..301M, 2003ApJ...587L..55C, 2003A&A...397..415C, 2004ApJ...617...81B, 2005ApJ...627L..93R, mw11}.
The effect of lensing by binary black holes has yet to be explored. Such systems are generated by the
merging of galaxies \citep{2002MNRAS.331..935Y, 2006ApJ...642L..21B, 2009ApJ...707L.184J}. 
Mergers between galaxies are observed and predicted in the hierarchical structure formation theory. 
The formation and evolution of binary black holes have been extensively studied using analytical and numerical methods 
(see, e.g.,  \citealt{2005LRR.....8....8M, 2009arXiv0906.4339C} for reviews and references therein). 
We briefly discuss the processes below.

When two galaxies merge, the orbits of their associated black holes will first decay through dynamical friction. 
The critical separation where the binary rotation velocity equals the
velocity dispersion of the host galaxy is often referred to in the
literature as the ``hardening radius'' of the binary (see eq. \ref{eq:hard}). This radius plays an important role in binary black hole evolution.
 At separations larger than the hardening radius, the black holes are
``dressed" with the inner cores of the stellar bulges belonging to their
original host galaxies. The binary separation shrinks below the hardening radius due to the
passage of individual stars which extract angular momentum from the
binary. These stars are removed from their orbits after the interaction
with the binary, leading to the appearance of a ``loss cone'' in phase
space. If this loss cone is not refilled, the orbital decay of the binary may stall, 
leading to a large population of galaxies with SMBBHs. 
The stalling radius is typically at several pc to 
several tens of pc 
(see, e.g., \citealt{2002MNRAS.331..935Y} and \citealt{2005LRR.....8....8M, 2009arXiv0906.4339C} for reviews). 

Some other (e.g., gas) processes need to bring the binary black holes closer, so that  
gravitational radiation can rapidly merge the binary black holes into a single one. 
The problem of whether loss cones are refilled fast enough is still unsolved. 
It is, therefore, unknown how many supermassive binary black holes there are in the universe. 
Thus, any probe of this population will provide additional constraints on the formation and evolution of supermassive binary black holes.


The purpose of this paper is to study the effects of SMBBHs on lensing properties using simple analytical models.
We do not consider the effects of the inner stellar cores associated with the black holes. In these simple models,
we show that the presence of SMBBHs can not only disturb the critical curves of
the primary lens galaxy but also create additional images.  
Many new lenses will be discovered with the next generation instruments, such as 
Pan-Starrs\footnote{http://pan-starrs.ifa.hawaii.edu/public/home.html} and LSST\footnote{http://www.lsst.org/lsst/scibook}.
Some of the more interesting cases will be observed with higher resolution and larger dynamical range using other instruments from which additional images, 
if detected, may provide direct evidence for the existence of SMBHs or SMBBHs in galaxies. 
However, these images are usually very faint and close to each other. The Square Kilometer Array (SKA\footnote{http://www.skatelescope.org/}),  
if equipped with a very long baseline, 
which will provide high angular resolution and dynamic range images (maps), may provide the best possibility.  
If lenses have shallow (non-singular) central profiles \citep{2004AJ....127.1917T}, central core images are predicted, but they
may be destroyed by the presence of SMBBHs \citep{2001MNRAS.323..301M, 2005ApJ...627L..93R}. We also touch upon this issue.

The outline of this paper is as follows. In section 2,  
we present the basic lensing model of a galaxy with SMBBHs,  e.i., the non-singular isothermal ellipsoid with SMBBHs.
The classification of SMBBHs is also discussed in section 2. 
In section 3, we focus on critical curves and caustics of a galaxy with SMBBHs.
In section 4, we discuss the cross sections of black hole images above a certain magnification 
threshold and estimate the probability of these images being detectable.
In section 5, we study the influence of SMBBHs on the core images in non-singular isothermal lens models.
Conclusions and a discussion are given in section 6. 
Throughout this paper we assume a flat $\Lambda$CDM cosmology with 
$\Omega_{\rm m,0} = 0.3$, $\Omega_{\Lambda,0} = 0.7$ and a Hubble constant $H_{0} = 100\, h~{\rm km~s^{-1}~Mpc^{-1}}$, $h = 0.7$.

\section{lens model with SMBBHs}

We model the lensing galaxy as a non-singular isothermal ellipsoid halo plus an SMBBH.
This model includes the singular isothermal ellipsoid (SIE) model as a special case. The SIE model is not only analytically tractable 
but also consistent with models of individual lenses, lens statistics, 
stellar dynamics and X-ray galaxies \citep{1989ARA&A..27...87F, 1993ApJ...416..425M, 
1995ApJ...445..559K, 1996ApJ...466..638K, 1996ApJ...464...92G, 1996ApJ...473..570G, 1997ApJ...488..702R, 
2002ApJ...575...87T, 2003ApJ...582...17K, 2003ApJ...595...29R, 2005ApJ...623..666R, 2006ApJ...649..599K, 2007ApJ...667..176G, 
2007ApJ...669...21P, 2008MNRAS.384..987C, 2008MNRAS.388..384D, 2009A&A...501..475T}. 
In other words, the core radii are expected to be small in elliptical galaxies \citep{2001ApJ...549L..33R, RT01, 2003ApJ...582...17K, 
2004Natur.427..613W, 2006ApJ...648...73B, 2007MNRAS.377.1623Z}. 
In $\S 2.1$ we outline the lensing basics 
for a non-singular isothermal ellipsoid plus a SMBBHs.
In $\S 2.2$, we focus on the classification of SMBBHs, which is important for understanding
their influence on gravitational lenses.  

\subsection{Non-singular Isothermal Lens Model with SMBBHs}

The dimensionless surface mass density distribution of a non-singular isothermal ellipsoid is given by
\begin{equation}
 \kappa = \frac{\Sigma}{\Sigmacr} =\frac{1}{2 q}\frac{1}{\sqrt{x_{1}^{2}+x_{2}^{2}/q^{2}+r_{\rm c}^{2}}},
\end{equation}
where $r_{\rm c}$ is the core radius, $q$ is the axis ratio, 
$\Sigmacr=c^{2} D_{\rm s}/(4 \pi G D_{\rm d} D_{\rm ds})$ is the critical surface density, 
$D_{\rm d}$, $D_{\rm s}$ are angular diameter distances from the observer to the lens and source, 
respectively, and $D_{\rm ds}$ is the angular diameter distance from the lens to the source. 
All the lengths $(x_{1}, x_{2}, r_{\rm c})$ are expressed in units of the critical radius, $\Rcr$, 
which is also called the Einstein radius. 
\begin{equation}
 \Rcr = D_{\rm d} \theta_{\rm E,SIS}, ~~~~~~~~~~~\theta_{\rm E,SIS}=4\pi\left( \frac{\sigma_{\rm v}}{c}\right)^2 \frac{D_{\rm ds}}{D_{\rm s}},  
\end{equation}
where $\theta_{\rm E,SIS}$ is the angle subtended by the critical radius on the sky
($\theta_{\rm E,SIS}\sim 0.2-3$ arcsec for typical lens galaxies), and the velocity dispersion $\sigma_{\rm v}$ is related to, 
but not necessarily identical to, the observable line-of sight velocity dispersion. 
We shall ignore this complication in our analysis and treat it as a parameter.
For purposes of illustration, the source is taken to be at redshift 2, and the lens is at redshift 0.5. 
The velocity dispersion is $\sigma_{\rm v} = 200~{\rm km\,s^{-1}}$, and axis ratio $q = 0.7$,  
which is the most probable axis ratio of early-type galaxies \citep{2007ApJ...658..884C}.

The lensing properties of the isothermal ellipsoid have been given by several authors (e.g., 
\citealt{1993ApJ...417..450K, 1994A&A...284..285K, 1998ApJ...495..157K}).
The lens equation including a SMBBHs is given by
\begin{align}
y_{1}&= x_{1}-\frac{\sqrt{q}}{\sqrt{1-q^{2}}} {\rm tan}^{-1}\left(\frac{\sqrt{1-q^{2}} x_{1}}{\Phi + r_{\rm c}/q}\right)   \nonumber \\
&-m_{1}\left( \frac{x_{1}-u_{1}}{r_{\rm a}^{2}}\right)-m_{2}\left( \frac{x_{1}-v_{1}}{r_{\rm b}^{2}}\right)~,   \\
y_{2}&= x_{2}-\frac{\sqrt{q}}{\sqrt{1-q^{2}}} {\rm tanh}^{-1}\left(\frac{\sqrt{1-q^{2}} x_{2}}{\Phi + r_{\rm c} q}\right)  \nonumber \\
&-m_{1}\left( \frac{x_{2}-u_{2}}{r_{\rm a}^{2}}\right)-m_{2}\left( \frac{x_{2}-v_{2}}{r_{\rm b}^{2}}\right)~, 
\end{align}
where $\Phi^{2} = q^{2}x_{1}^{2} + x_{2}^{2} + r_{\rm c}^{2}$, and
$m_{1}$, $m_{2}$ are the dimensionless mass of the two BHs respectively. 
We label the two BHs of SMBBHs as `$a$' and `$b$'. $r_{\rm a}$, $r_{\rm b}$ are the dimensionless projected distances 
from the images to the black hole `$a$', `$b$' on the lens plane. $(u_{1}, u_{2})$, $(v_{1}, v_{2})$ 
are the coordinates of `$a$' and `$b$' on the lens plane
\begin{align}
  r&=\sqrt{x_{1}^{2}+x_{2}^{2}}~,    \\
  r_{\rm a}&=\sqrt{(x_{1}-u_{1})^{2}+(x_{2}-u_{2})^{2}}~,     \\
  r_{\rm b}&=\sqrt{(x_{1}-v_{1})^{2}+(x_{2}-v_{2})^{2}}~.
\end{align}
We also define  
\begin{equation}
m = m_{1} + m_{2}, 
\end{equation}
\begin{equation}
 m = \frac{M_{\rm bh}}{M_{\rm cr}}, ~~~~~~~~~~~~~~~M_{\rm cr} = \frac{\pi \sigma_{\rm v}^{2} \Rcr}{G},  \label{eq:m}
\end{equation}
where $M_{\rm bh}$ is the total mass of the supermassive binary black holes. Physically, $M_{\rm cr}$ is the mass of the 
galaxy contained within a cylinder with radius $\Rcr$, hence $m$ is 
the ratio of the total mass of the SMBBHs to the projected mass of the galaxy within $\Rcr$.
We assume that the correlation of the total mass of SMBBHs and velocity dispersion is the same as that for a single black hole, which 
has been studied by many authors 
\citep{2000ApJ...539L...9F, 2000ApJ...539L..13G, 2002ApJ...574..740T, 
2003MNRAS.342..501N, 2006ApJ...641...90R, 2008MNRAS.386.2242H, 2009ApJ...698..198G}. 
In our paper, we use the recent correlation found  by \cite{2009ApJ...698..198G}
\begin{equation}
M_{\rm bh}\approx 10^{8.23} M_{\odot} \left({\sigma_{\rm v} / 200\kms}\right)^{3.96}~, 
\end{equation}
for elliptical galaxies, which dominate the lensing cross-sections.
The dimensionless total mass of the  SMBBHs is thus given by 
\begin{equation}
 m = 2.5 \times 10^{-3} h\left( \frac{\sigma_{\rm v}}{200\kms}\right)^{-0.04}.
\end{equation}
Notice that $m$ has little dependence on the velocity dispersion, although we caution that there is substantial scatter around this correlation.
The magnification ($\mu$) is given by
\begin{equation}
 \mu^{-1}= \frac{\partial y_{1}}{\partial x_{1}}\frac{\partial y_{2}}{\partial x_{2}}
-\frac{\partial y_{1}}{\partial x_{2}}\frac{\partial y_{2}}{\partial x_{1}}.
\label{mu_1} 
\end{equation}
Throughout this paper, we quote the absolute magnification only (without parity).

Including the source position, the lens model has 11 degrees of freedom ($\vec{y}, \vec{u}, \vec{v}, m_{1}/m_{2}, \sigma_{\rm v}, q, r_{\rm c}$),
 even if we ignore the scatter in the correlation between $m$ and$\sigma_{\rm v}$.
 The parameter space is large, and thus in this paper we limit ourselves to illustrative examples. Also 
 notice that if we set the core size $r_{\rm c} = 0$, 
the non-singular isothermal model becomes a singular isothermal ellipsoid model, for which the influence of SMBBHs  
will be studied in $\S3$ and $\S4$. 

The lens equation has to be solved numerically to yield the image positions; their magnifications can then be found through Eq.~(\ref{mu_1}). 
We can, however, study the influence of SMBBHs on gravitational lenses using critical curves and caustics.
Critical curves are image positions whose magnifications are infinite 
($\mu^{-1} = 0$). They map into caustics in the source plane. Caustics mark discontinuities in the number of images,  
so we can study the differences in critical curves and caustics, between lens models with and without 
SMBBHs to understand their lensing effects (see  $\S 3$). In order to see whether additional images created by black holes are observable, 
we examine the cross section when these images have magnifications above a
certain threshold, e.g. $\mu_{\rm th} > 10^{-3}, 10^{-4}$, and $10^{-5}$ (see \S 4 and \S 5).

\subsection{Classification of SMBBHs} 

We classify each SMBBH in terms of the separation of its two members.
First, we calculate the condition that the rotation velocity of supermassive binary black holes is equal to 
the velocity dispersion. In this case, the separation of the supermassive binary black holes is called the ``hardening" radius ($\dcr$),
\begin{equation}
 \frac{GM_{\rm bh}}{4\dcr} = v^{2} \approx 2\sigma_{\rm v}^{2},
\end{equation}
\begin{equation}
 \dcr=\frac{GmM_{\rm cr}}{8\sigma_{\rm v}^{2}}, \label{eq:hard}
\end{equation}
where $M_{\rm cr}$ and $m$ are defined in eq. (\ref{eq:m}).
The hardening radius $\dcr$ can be estimated as
\begin{equation}
 \dcr \approx 3.53~{\rm pc} \left(\frac{\sigma_{\rm v}}{200\kms} \right)^{1.96} .
\end{equation}
If the separation is smaller than this radius, we call these SMBBHs `hard', 
otherwise `soft'. As discussed in the introduction, 
this may also be the most probable separation of SMBBHs 
(${\rm several~pc} \sim {\rm several~10pc}$, \citealt{2002MNRAS.331..935Y}) if the ``loss cone" is not refilled.

The ratio of the ``hardening" radius to the Einstein radius of its host galaxy is
\begin{equation}
 \frac{\dcr}{\Rcr}
=\frac{\pi m}{8} \approx 6.87\times10^{-4} \left(\frac{\sigma_{\rm v}}{200\kms} \right)^{-0.04} . \label{eq:units}
\end{equation}

Two other ratios are relevant: the Einstein radius of BH to the Einstein radius of its host galaxy, and 
the ``hardening" radius to the Einstein radius of the BH's.  Again assuming a lens redshift of 0.5 and a source redshift of 2, the first ratio is given by
\begin{equation}
 \frac{\theta_{\rm E,BH}}{\theta_{\rm E,SIS}} = \sqrt{m} \approx 4.18 \times10^{-2} \left(\frac{\sigma_{\rm v}}{200\kms} \right)^{-0.02},
\end{equation}
and the second by
\begin{equation}
 \frac{\theta_{\rm BH}}{\theta_{\rm E,BH}} =\frac{\pi \sqrt{m}}{8} 
 \approx 1.64 \times10^{-2} \left(\frac{\sigma_{\rm v}}{200\kms} \right)^{-0.02}.
\end{equation}
Both quantities are on the order of a few percent.

\section{Critical Curves and Caustics of Singular Isothermal Ellipsoid lens with SMBBHs}

As discussed in the previous section, 
critical curves represent the positions of images of infinite magnification, while  
caustics mark discontinuities in the number of images. For singular potentials, there may also exist so-called pseudo-caustics: 
when a source crosses pseudo-caustic, the image number changes by one, rather than by two (\citealt{ew98}).
In this section, we study how the SMBBHs affect the critical curves and caustics of 
singular isothermal gravitational lenses (i.e., $r_{\rm c} = 0$).

\begin{figure*}
\begin{minipage}{105mm}
\centering
\includegraphics[angle=-90,width=1.0\columnwidth]{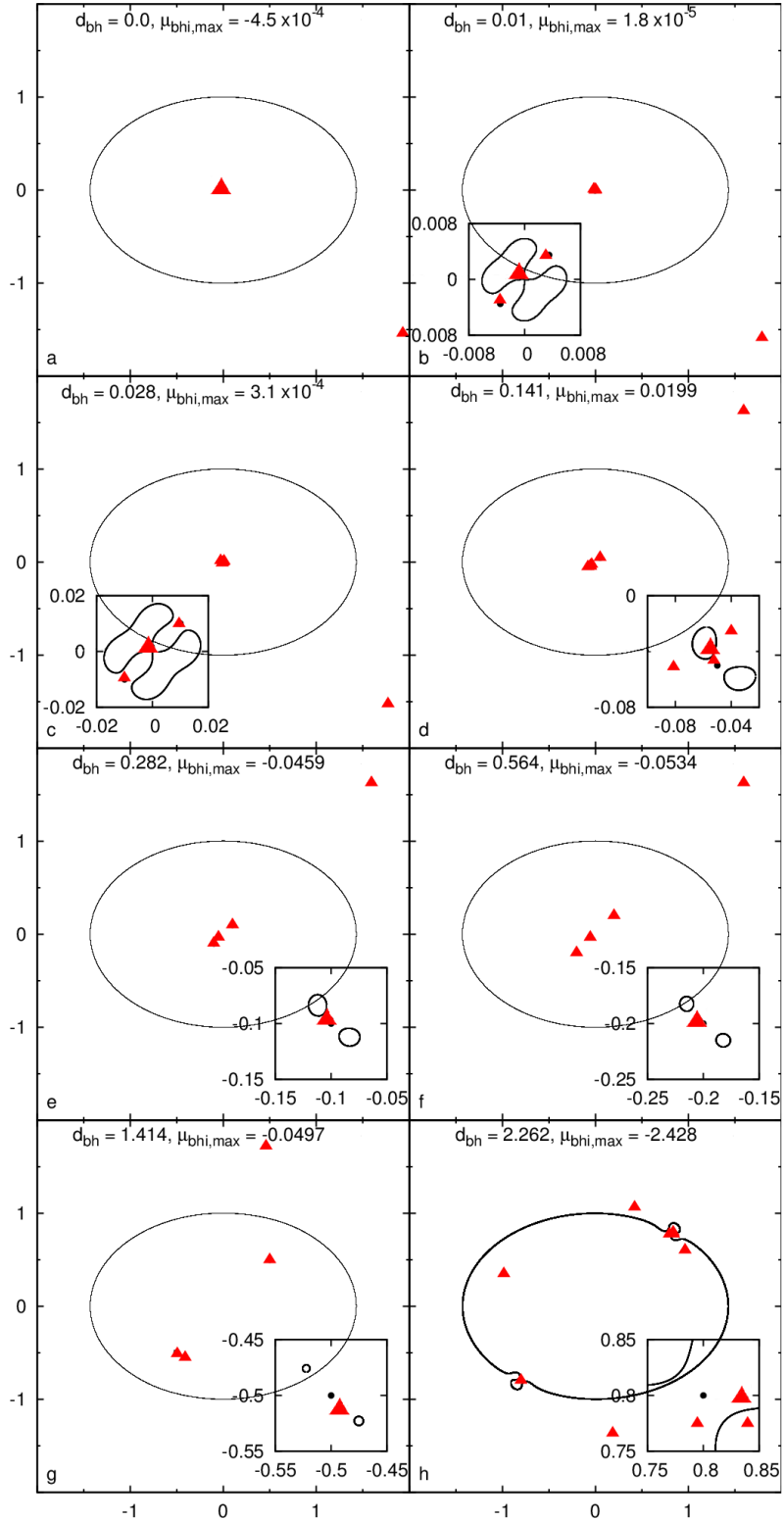}
\caption{Critical curves and examples of images  for the case of a SMBBH in 
a singular isothermal ellipsoid. The mass ratio for the binary system is 
$m_{1}/m_{2} = 1$, and the separation $d_{\rm bh}$ is indicated in each panel.
$d_{\rm bh}$ is in units of $r_{\rm E,SIS}$,  
the Einstein radius of the singular isothermal sphere model.
The black curves show the critical curves,  
red triangles show the image positions.
We also plot the positions of the SMBBHs as black points. 
Each case has $m = 2.5\times10^{-3} h$,  
and the galaxy is modelled as a singular isothermal ellipsoid with velocity dispersion $\sigma_{\rm v} = 200~{\rm km~s^{-1}}$ and axis ratio $q = 0.7$.
The black hole image with the highest magnification is indicated as a larger triangle for each case.} 
\label{lens_c_cc}
\end{minipage}
\end{figure*}

\begin{figure*}
\begin{minipage}{105mm}
\includegraphics[angle=-90,width=1.0\columnwidth]{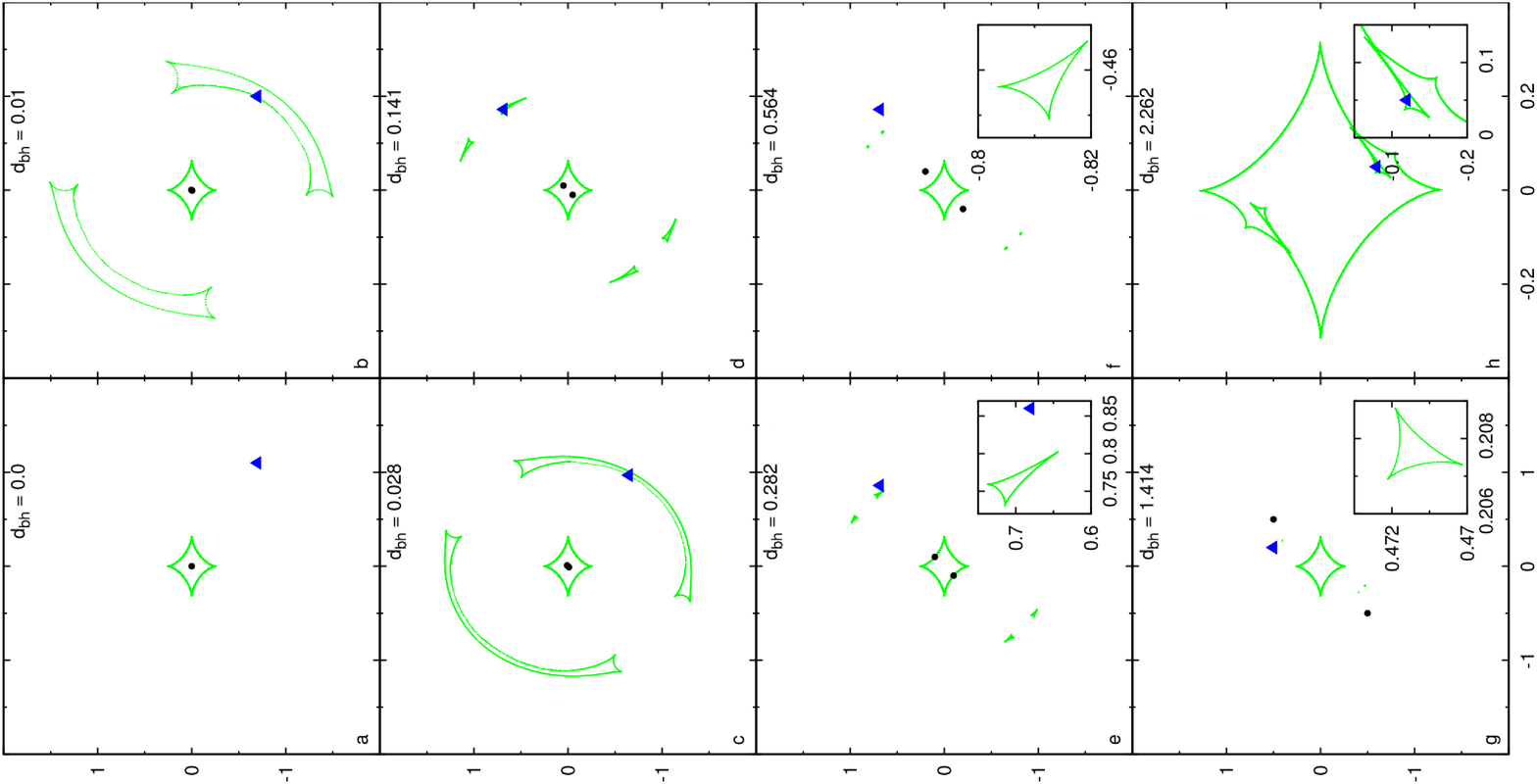}
\caption{Caustics for the case of a  SMBBH in 
a singular isothermal galaxy for the same parameters as in Fig.  \ref{lens_c_cc}. Some example source positions are shown with corresponding images shown in Fig. \ref{lens_c_cc}. 
The green curves show the caustics, while the
blue triangles show positions of the sources.
We also plot the positions of the SMBBHs in these panels (black points). 
All SMBBHs cases have $m = 2.5\times10^{-3} h$,  
and the galaxy is modelled as a singular isothermal ellipsoid with velocity dispersion $\sigma_{\rm v} = 200~{\rm km~s^{-1}}$ and axis ratio $q = 0.7$.} 
\label{source_c_cc}
\end{minipage}
\end{figure*}

We show the critical curves and caustics of a singular isothermal ellipsoid model 
with central SMBBHs at different separations in Figs.~\ref{lens_c_cc} and \ref{source_c_cc} respectively. 
The separation ($d_{\rm bh}$) is always expressed in units of the Einstein radius, which corresponds to about $5.1$\,kpc for our example. As can be seen, there are some critical curves near the BHs (which for convenience we will call them black hole critical curves) 
when the separation of SMBBHs is $ 0<\dbh \le 1.6$. 
The black hole critical curves are a single continuous curve when $ \dbh \lesssim 0.05$, 
while they become several disjoint curves when $\dbh \ge 0.05$ (Fig.~\ref{lens_c_cc}). 
These behaviours are also reflected in the caustics (Fig.~\ref{source_c_cc}). 
When the separation becomes even larger ($\dbh \approx 0.5$), the black hole critical curves become smaller and smaller.
However, when the separation of SMBBHs becomes very large ($\dbh \gtrsim 1.6$),
the critical curve of the primary lens can be disturbed by BHs 
and the black hole critical curves are merging into the primary critical curves, 
so we have peculiar caustics as shown in the bottom right panel of Fig.~\ref{source_c_cc}. 
Such large separations are not expected to be common in SMBBHs 
\citep{2002MNRAS.331..935Y,2009arXiv0906.4339C}.

We also show some examples of images in Fig.~\ref{lens_c_cc} and the positions of corresponding sources in Fig.~\ref{source_c_cc}, 
we can see that each black hole of SMBBHs can generate BH-images near 
itself (Fig.~\ref{lens_c_cc}), if the source is located in special positions (Fig.~\ref{source_c_cc}), 
e.g. near the pseudo-caustics, there can be multiple BH-images for 
each black hole.
When the separation of the SMBBHs is zero (i.e., a single BH),
and the source is outside the primary elliptical critical curve (also the pseudo-caustic, see  
 panel (a) of Fig.~\ref{source_c_cc}), 
there is a BH-image near the SMBBHs, which is shown in panel (a) of Fig.~\ref{lens_c_cc}. 
As is well known for a singular isothermal ellipsoid lens, if the
source is located outside the pseudo-caustic there are no
multiple-images, so the central image must be generated by the SMBH.
When the separations are not large, as illustrated in panels (b) and
(c) of Fig.~\ref{lens_c_cc}, the black hole critical curve is
continuous, and we find that there are three BH images for these two
cases of source positions, as shown in panels (b) and (c) of
Fig.~\ref{source_c_cc}.  In panel (d) of Fig.~\ref{lens_c_cc}, the
SMBBH separation is larger than 0.05 and the black hole critical
curves as well as the caustics become disjoint (see panel (d) of
Fig.~\ref{source_c_cc}).  The black hole close to the source generates
one BH-image, while the black hole farther away from the source
generates three BH-images. There is also another image around the position
$(x=-0.081, y=-0.051)$ associated with the SIE lens.  The cases in
panels (e), (f), and (g) of Fig.~\ref{lens_c_cc} are similar:
BH images are generated close to each BH in addition to a macro image
due to the SIE lens close to the primary critical curve. The most
unusual case is shown in the bottom right panels of
Fig.~\ref{lens_c_cc} and Fig.~\ref{source_c_cc}.  The separation is
such large that the two black holes distort the primary critical
curve, 8 images are created in total, including 4 BH-images
(see panel (h) of Fig.~\ref{lens_c_cc}), when the source is
located in a special position (see panel (h) of
Fig.~\ref{source_c_cc}).  Some of these BH-images are bright ($|\mu|
> 0.425$) enough to be potentially detectable, but the
probability may be low because a large separation ($11.6$ kpc for our
illustrative example) is required, and most black holes may not be at
such large distances.
    
In the other extreme, we have very hard SMBBHs (i.e., with very small separations, 
$d \lesssim 10^{-6}$ in the Einstein radius). In principle, the rapid rotation of the binary black hole may lead to variations in the magnification. 
However, for this to be observable, its timescale needs to be relatively short, 
$T/4 \lesssim 10\,{\rm yr}$, where the period $T=(d^3/(GM_{\rm bh}))^{1/2}$. 
This requires a separation $d\lesssim 0.04$pc for a total black hole mass of 
$1.7\times 10^8M_\odot$ corresponding to $\sigma_{\rm v}=200$\,km/s.  This separation is even smaller than $\dcr$, 
and thus the binary black holes will essentially appear as a single one for lensing purposes. We conclude that in general
the binary rotation effect will be difficult to detect using current or even future facilities.


\section{Cross Sections and Probability}

In this section, we investigate BH-images with magnifications above several thresholds: 
$\muth=10^{-3}, 10^{-4}$ and $10^{-5}$; 
such images are potentially detectable (see the discussion). 
We calculate cross-sections in the singular isothermal ellipsoid model for producing at least one or two such BH-images.

The cross-sections are calculated by constructing the magnification map 
(in the image plane) and identifying the region where the magnification is greater than some minimum value $\mu_{\rm th}$. 
We then map this region onto the source plane, from which we calculate the source cross-section.
The cross section in the case of having at least two BH-images above a magnification threshold reflects the overlapping 
region where each member of the SMBBHs satisfies the condition. 
To do this, we first calculate the cross section for the single BH case and then for binary BHs. 
The overlapping cross-section is obtained by subtracting the cross-section of the binary BHs 
from the total cross-section of the two single black holes.

\begin{figure}
\includegraphics[width=1.0\columnwidth]{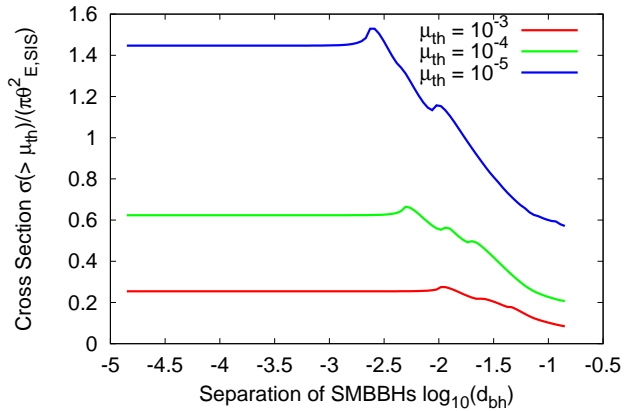}
\caption{Cross-section of BH-images of SIE-SMBBHs model as a function of separation $d_{\rm bh}$ of 
SMBBHs with different thresholds of magnification.
The red, green and blue curves correspond to  $\muth=10^{-3}, 10^{-4}$ and $10^{-5}$ (from bottom to top) respectively. 
All cases have $m = 2.5 \times 10^{-3} h$, and $m_{1}/m_{2} = 1$. 
Each galaxy is modelled as a singular isothermal ellipsoid with velocity dispersion
 $\sigma_{\rm v} = 200~{\rm km~s^{-1}}$, and axis ratio $q = 0.7$. 
$\mu_{\rm th}$ is the threshold of the magnification for BH images.}   
\label{sie_cs_dbh}
\end{figure}

Fig~\ref{sie_cs_dbh} shows the cross-sections as a function of the separation between the binary black holes for three thresholds
$\muth=10^{-3}, 10^{-4}$ and $10^{-5}$. As expected, the higher the threshold, the lower the cross section of the BH-images. 
The peak of the cross-section also moves to a larger separation when the threshold increases.

\begin{figure}
\includegraphics[width=1.0\columnwidth]{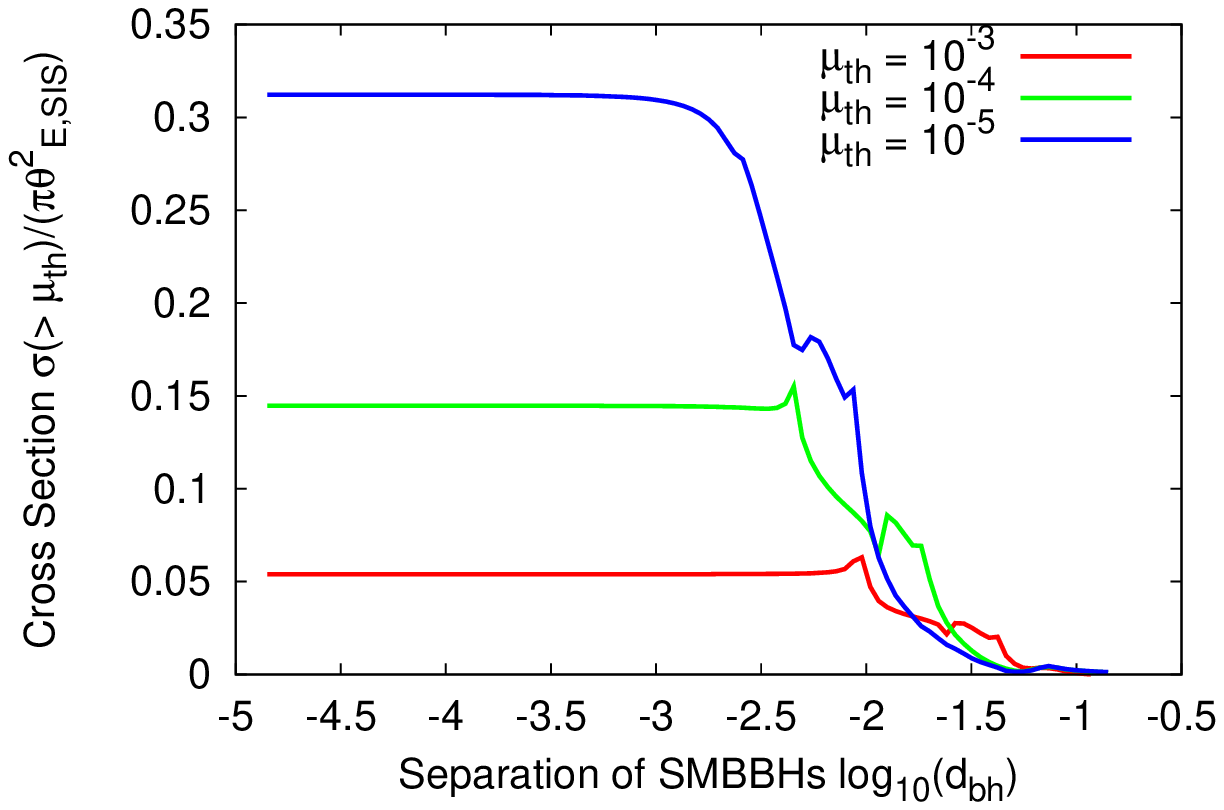}
\caption{Similar to Fig.~\ref{sie_cs_dbh}, except that cross section is for the case when  
both BHs create BH-images with magnification greater than three thresholds as in Fig.~\ref{sie_cs_dbh}.  
All SMBBHs cases have $m = 2.5 \times 10^{-3} h$, and the mass ratio $m_{1}/m_{2} = 1$. 
Each galaxy is modelled as singular isothermal ellipsoids 
with velocity dispersion $\sigma_{\rm v} = 200~{\rm km~s^{-1}}$ and axis ratio $q = 0.7$. 
$\mu_{\rm th}$ is the threshold of the magnification.}
\label{sie_cs_dbh_both}
\end{figure}

Fig.~\ref{sie_cs_dbh_both} illustrates 
the cross-section that we can detect at least 2 BH-images, which are generated by both members of the SMBBHs respectively. 
As shown in Fig.~\ref{sie_cs_dbh_both}, only when the separation is small enough,  two BH-images can be generated, 
and the higher the $\muth$ is, the lower the cross section of BH-images is, as expected.



\section{Suppression of core images in a non-singular isothermal galaxy with SMBBHs}

In this section, 
we investigate how core images of the non-singular isothermal ellipsoid (NIE) lens model are affected by central SMBBHs.
As is well known, non-singular isothermal model can generate a faint core image, 
and the magnification and position of the core images are sensitive to the core size. 
Observationally,  very few central images have been observed (for more, see the Discussion), 
which can be used to put an upper limit on the core radius 
(e.g., \citealt{2001ApJ...549L..33R, RT01, 2003ApJ...582...17K, 2004Natur.427..613W, 2006ApJ...648...73B, 2007MNRAS.377.1623Z}).  
Not surprisingly, as in the case of a single central black hole, 
the presence of SMBBHs can also demagnify and suppress the observability of central core images.

\begin{figure}
\includegraphics[width=1.0\columnwidth]{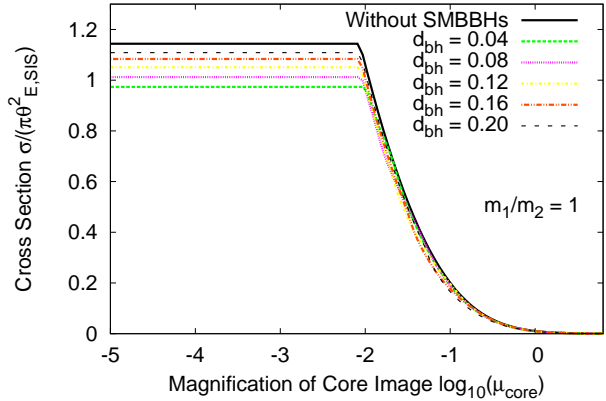}
\includegraphics[width=1.0\columnwidth]{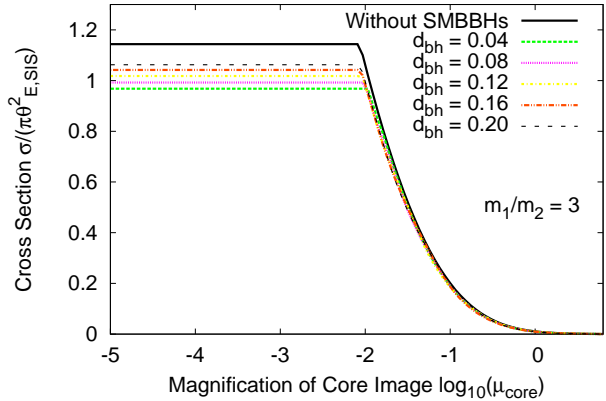}
\caption{Cumulative distribution functions for the magnification of core-images ($\mu_{core}$) 
in the non-singular isothermal halo with SMBBHs.
The heavy lines show the case without BHs. The light lines show SMBBHs cases with separations $\dbh = 0.04, 0.08, ..., 0.20$. 
All SMBBHs cases have $m = 2.5 \times 10^{-3} h$, and $m_{1}/m_{2} = 1$ (top),  $m_{1}/m_{2} = 3$ (bottom).
Each galaxy is modelled as a non-singular isothermal ellipsoid with core radius $r_{\rm c} = 0.05$, 
velocity dispersion $\sigma_{\rm v} = 200~{\rm km~s^{-1}}$ and axis ratio $q = 0.7$.}
\label{cis_cs_mu_ne_core}
\end{figure}

\begin{figure}
\includegraphics[width=1.0\columnwidth]{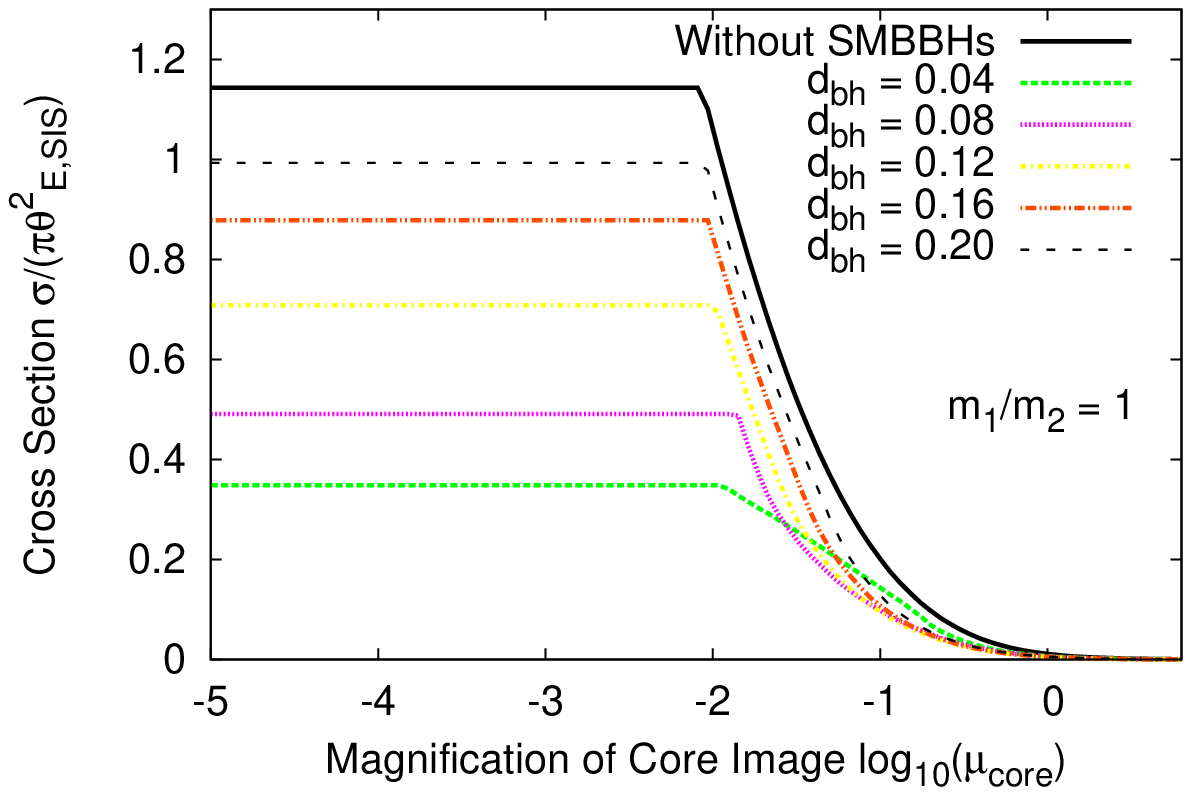}
\includegraphics[width=1.0\columnwidth]{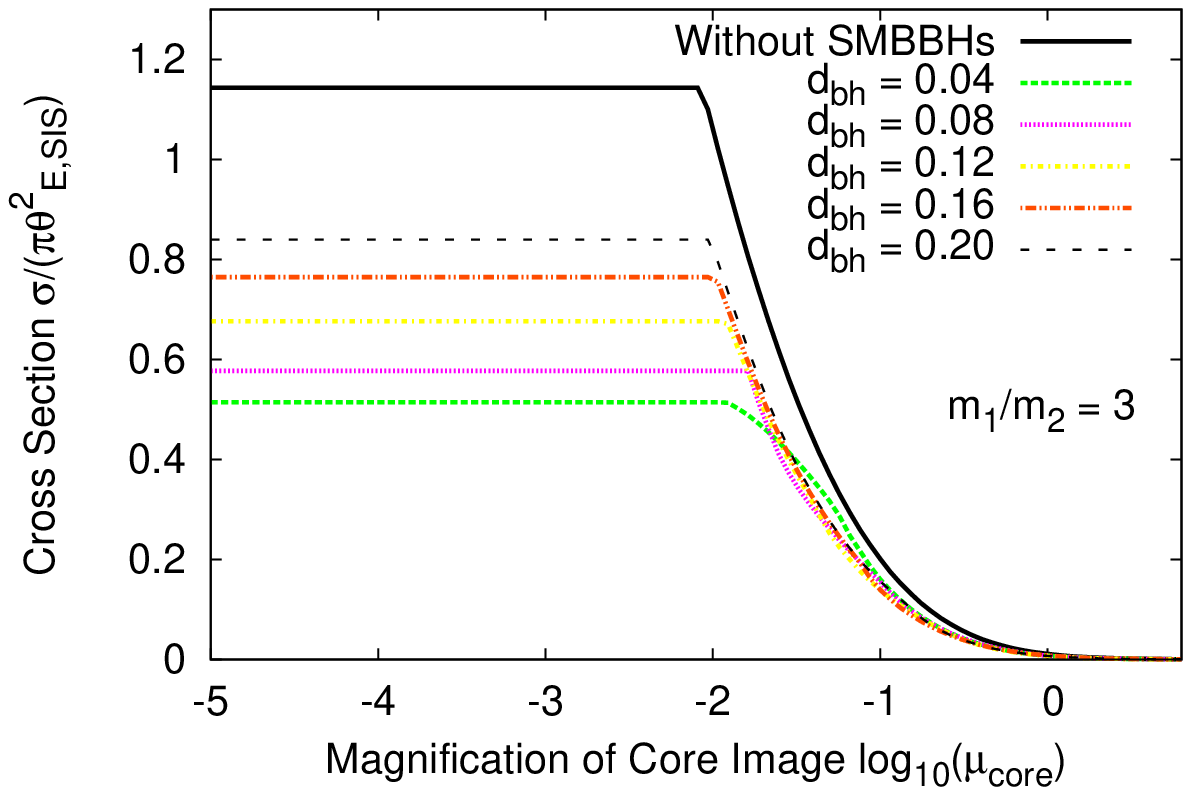}
\caption{Similar to Fig.~\ref{cis_cs_mu_ne_core}, except that the total dimensionless mass of the SMBBHs is $m = 0.01$.}
\label{cis_cs_mu_ne_core2}
\end{figure}

We again set the velocity dispersion $\sigma_{\rm v}$ equal to $200~{\rm km~s^{-1}}$, 
the axis ratio $q$ equal to $0.7$, and adopt a core size $r_{\rm c} = 0.05$. 
Fig.~\ref{cis_cs_mu_ne_core} shows the cumulative distribution function for the magnification of core images ($\mu_{\rm core}$). 
SMBBHs suppress the faint end of the distribution, leaving the bright end largely unaffected. 
A smaller separation will suppress the faint end of the distribution more effectively 
than a larger one. For $d_{\rm bh} \lesssim 0.04$, the suppression is on the order of $12\%$, 
and for $d_{\rm bh} = 0.20$ the suppression is on the order of $3\%$.
Non-equal mass SMBBHs lead to smaller variations between different separations 
than equal mass systems. For example, for $d_{\rm bh} = 0.20$ with $m_1/m_2=3$, the suppression is on the order of $7\%$, 

Fig.~\ref{cis_cs_mu_ne_core2} shows the cumulative distribution function for the magnification of core images 
in non-singular isothermal lens with more massive SMBBHs ($m=0.01$). 
As expected, the cross section of core images are suppressed more at the faint end of the distribution.
To summarize, if double black holes exist in multiply-imaged systems, then they will lead to small changes in
the constraints on the central mass distributions in lenses.

\section{Discussion and Conclusions}

In this paper, we have studied the lensing configuration due to supermassive binary black holes (SMBBHs). 
We show typical examples of critical curves, caustics, and image configurations. 
Similar to a single black hole, SMBBHs can create additional images close to them. 
While we have adopted illustrative values for the axis ratio and orientation, we have also explored other values and found 
no significant dependence on these parameters.

For BH-images to be observable, they have to be bright enough to be detected. 
We have examined the cross-sections for producing BH-images with magnification above the thresholds, 
$\mu_{\rm th}=10^{-3}, 10^{-4}$ and $10^{-5}$ relative to that of producing multiple-images, 
$\pi \theta^2_{\rm E, SIS}$. We write this ratio as ${\cal R}_{\rm BH}$.
The probability of being able to detect BH-images in multiply-imaged systems is given by 
$P_{\rm BH}(> \mu)~\approx~{\cal R}_{\rm BH}(> \mu)~\fBH$, where
$\fBH$ is the fraction of galaxies with SMBBHs.

The values of ${\cal R}_{\rm BH}$ can be read off from 
Fig.~\ref{sie_cs_dbh}. 
Thus,
the observational probability of the BH-images in a multiply-imaged system for $\muth=10^{-3},10^{-4}$ and $10^{-5}$, are about $0.2\fBH$, 
$0.6\fBH$ and $1.4\fBH$ respectively. The probabilities, where both BHs generate at least one BH-image for the same thresholds, are
$0.05\fBH$, $0.15\fBH$ and $0.3\fBH$ respectively, i.e., approximately a factor of four lower.

As shown in Fig.~\ref{lens_c_cc}, BH-images are very close to the BHs, 
so the separation between the brighter BH-images is approximately the separation of the SMBBH, 
of which the most probable value is $\sim 10^{-4}$ in units of $R_{\rm E,SIS}$ (see eq.  \ref{eq:units}). 
Thus the resolution must be better than $\approx 10^{-4}$ arcsec, which is already achievable by VLBI techniques. 
On the other hand, we need very large dynamic range to detect the 
BH-images. When the magnification of BH-image is $10^{-3}$, the dynamic range
required is $\mu_{\rm max}/10^{-3}$, where $\mu_{\rm max}$ is the magnification of the brightest image. 
For most cases, the largest magnification $\mu_{\rm max}$ is about a few to ten. 
Thus if we want to detect a BH-image whose magnification is $10^{-3}$, as a 
conservative estimate, the dynamic range needs to be $\gtrsim 10^{4}$.

It is interesting to speculate what we can learn if we do observe two BH-images. 
We have 8 constraints (from the positions of two black hole images and 
two macro-images produced by the singular isothermal ellipsoid lens) 
while we have at least 10 parameters even if the core size is taken to be zero. 
Clearly the system is under-constrained. 
%
%
The number of parameters will be even larger in more complicated models, 
so we will not be able to determine the parameters uniquely. 
Furthermore, multiple BH images can also be produced by a single off-centre black hole \citep{mw11}, which may complicate the interpretation.

 
The presence of SMBBHs can suppress the faint end of the cumulative
distribution for the magnification of core images, while leaving the
bright end largely unaffected. Their effects will need to be accounted
for in the constraint on the central mass profiles (e.g., core
radius).  There are presently two known lenses with a core
image (PMN J1632$-$0033, \citealt{2004Natur.427..613W} and SDSS
J1004$+$4112, \citealt{2008PASJ...60L..27I}). It is not surprising
that SDSS J1004$+$4112 has a core image, because it is a cluster lens
with a shallow NFW \citep{1997ApJ...490..493N} profile. For PMN
J1632$-$0033, the evidence for the core image has been discussed in detail by
\citet{2004Natur.427..613W}.  At frequencies higher than 1.7 GHZ, 
the logarithmic slopes of flux density ratio vs. 
frequency for the three images are entirely consistent with the third image (C) being the elusive and long sought-after central image. 
At 1.7 GHZ, image C is fainter than expected, 
which may be due to absorption and scintillation through the dense lens galaxy centre. 
In addition, they predict a fourth image induced by the central BH at $<10\%$ level of the flux of image C, which, 
if detected with VLBI techniques, will provide a measurement of the BH mass (\citealt{2004Natur.427..613W}).

To summarize, gravitational lensing can in principle be used to detect single BHs and
SMBBHs in galaxies through the extra images they create. However,
these images are usually very faint and close to each other, so they
pose challenges for current instruments both in terms of resolution and sensitivity. For example, 
VLBI techniques may have sufficient resolution, but the dynamical range achievable currently may be insufficient to detect multiple, faint BH images. 
Another complication may arise because of the confusion of central images with radio emission from
the lens galaxy. However, most lensing galaxies are ellipticals and so their central AGNs may be weak. 
Even so, at very high sensitivity/dynamical range, confusion with central AGNs may still be an issue. As discussed by 
\citet{2004Natur.427..613W} and \citet{2005ApJ...627L..93R}, 
we can use the usual tests - a common spectrum (or flux density ratio vs. frequency), 
surface brightness and correlated variability (time delays) - to differentiate central images from an AGNs in  the lens galaxy. 
For example, for B2108$+$213 \citep{2005MNRAS.356.1009M}, \citet{2008MNRAS.384.1701M} compared the spectrum of the central radio source with those 
of lensed images and concluded that the central source is an AGN rather than a lensed image.
Complications due to absorption and scintillation can be overcome by observing at high frequencies in the radio 
(since their effects scale as $\nu^{-2}$). Furthermore, if multiple BH images are discovered, 
then the confusion may be less of an issue since there is likely only one central AGN.

Future surveys using optical telescopes, such as Pan-Starrs and LSST, 
can provide a much larger sample of lenses, which may be used to identify particularly promising cases for further studies.
Future generation of instruments, in particular the Square Kilometer Array, will provide 
very high-contrast ($\gtrsim 10^{6}$) and high-resolution ($\lesssim 10^{-4}$ arcsec) imaging capabilities.
We remain cautiously optimistic that binary black holes can be 
independently discovered through careful observations of multiply-imaged systems, especially in the radio.

\section*{Acknowledgments}
We would like to thank Chuck Keeton, Youjun Lu, Xinzhong Er and Dandan
Xu for advice and useful discussions on the topic. We also acknowledge
an anonymous referee for criticisms that improved the paper. 
This research was supported by the Chinese Academy of Sciences (SM). 
NL and LG acknowledges support from the one-hundred-talents program of the
Chinese academy of science(CAS), the National basic research program
of China (973 program under grant No. 2009CB24901), the {\small NSFC}
grants program (No. 10973018), the Partner Group program of the Max
Planck Society and an STFC Advanced Fellowship. This work was supported 
in part by NSF grant AST-0907890 and NASA grants NNX08AL43G and NNA09DB30A (AL). 
\bibliographystyle{mn2e}
\bibliography{./lens_BBHs}

\begin{thebibliography}{}

\bibitem[\protect\citeauthoryear{{Bandara}, {Crampton} \& {Simard}}{{Bandara}
  et~al.}{2009}]{2009ApJ...704.1135B}
{Bandara} K.,  {Crampton} D.,    {Simard} L.,  2009, \apj, 704, 1135

\bibitem[\protect\citeauthoryear{{Berczik}, {Merritt}, {Spurzem} \&
  {Bischof}}{{Berczik} et~al.}{2006}]{2006ApJ...642L..21B}
{Berczik} P.,  {Merritt} D.,  {Spurzem} R.,    {Bischof} H.-P.,  2006, \apjl,
  642, L21

\bibitem[\protect\citeauthoryear{{Bowman}, {Hewitt} \& {Kiger}}{{Bowman}
  et~al.}{2004}]{2004ApJ...617...81B}
{Bowman} J.~D.,  {Hewitt} J.~N.,    {Kiger} J.~R.,  2004, \apj, 617, 81

\bibitem[\protect\citeauthoryear{{Boyce}, {Winn}, {Hewitt} \& {Myers}}{{Boyce}
  et~al.}{2006}]{2006ApJ...648...73B}
{Boyce} E.~R.,  {Winn} J.~N.,  {Hewitt} J.~N.,    {Myers} S.~T.,  2006, \apj,
  648, 73

\bibitem[\protect\citeauthoryear{{Chen}}{{Chen}}{2003a}]{2003ApJ...587L..55C}
{Chen} D.,  2003a, \apjl, 587, L55

\bibitem[\protect\citeauthoryear{{Chen}}{{Chen}}{2003b}]{2003A&A...397..415C}
{Chen} D.,  2003b, \aap, 397, 415

\bibitem[\protect\citeauthoryear{{Choi}, {Park} \& {Vogeley}}{{Choi}
  et~al.}{2007}]{2007ApJ...658..884C}
{Choi} Y.,  {Park} C.,    {Vogeley} M.~S.,  2007, \apj, 658, 884

\bibitem[\protect\citeauthoryear{{Colpi} \& {Dotti}}{{Colpi} \&
  {Dotti}}{2009}]{2009arXiv0906.4339C}
{Colpi} M.,  {Dotti} M.,  2009, ArXiv e-prints

\bibitem[\protect\citeauthoryear{{Czoske}, {Barnab{\`e}}, {Koopmans}, {Treu} \&
  {Bolton}}{{Czoske} et~al.}{2008}]{2008MNRAS.384..987C}
{Czoske} O.,  {Barnab{\`e}} M.,  {Koopmans} L.~V.~E.,  {Treu} T.,    {Bolton}
  A.~S.,  2008, \mnras, 384, 987

\bibitem[\protect\citeauthoryear{{Di Matteo}, {Croft}, {Springel} \&
  {Hernquist}}{{Di Matteo} et~al.}{2003}]{2003ApJ...593...56D}
{Di Matteo} T.,  {Croft} R.~A.~C.,  {Springel} V.,    {Hernquist} L.,  2003,
  \apj, 593, 56

\bibitem[\protect\citeauthoryear{{Dye}, {Evans}, {Belokurov}, {Warren} \&
  {Hewett}}{{Dye} et~al.}{2008}]{2008MNRAS.388..384D}
{Dye} S.,  {Evans} N.~W.,  {Belokurov} V.,  {Warren} S.~J.,    {Hewett} P.,
  2008, \mnras, 388, 384

\bibitem[\protect\citeauthoryear{{Evans} \& {Wilkinson}}{{Evans} \&
  {Wilkinson}}{1998}]{ew98}
{Evans} N.~W.,  {Wilkinson} M.~I.,  1998, \mnras, 296, 800

\bibitem[\protect\citeauthoryear{{Fabbiano}}{{Fabbiano}}{1989}]{1989ARA&A..27...87F}
{Fabbiano} G.,  1989, \araa, 27, 87

\bibitem[\protect\citeauthoryear{{Feoli}, {Mancini}, {Marulli} \& {van den
  Bergh}}{{Feoli} et~al.}{2011}]{2011GReGr..43.1007F}
{Feoli} A.,  {Mancini} L.,  {Marulli} F.,    {van den Bergh} S.,  2011, General
  Relativity and Gravitation, 43, 1007

\bibitem[\protect\citeauthoryear{{Ferrarese}, {C{\^o}t{\'e}}, {Dalla
  Bont{\`a}}, {Peng}, {Merritt}, {Jord{\'a}n}, {Blakeslee}, {Ha{\c s}egan},
  {Mei}, {Piatek}, {Tonry} \& {West}}{{Ferrarese}
  et~al.}{2006}]{2006ApJ...644L..21F}
{Ferrarese} L.,  {C{\^o}t{\'e}} P.,  {Dalla Bont{\`a}} E.,  {Peng} E.~W.,
  {Merritt} D.,  {Jord{\'a}n} A.,  {Blakeslee} J.~P.,  {Ha{\c s}egan} M.,
  {Mei} S.,  {Piatek} S.,  {Tonry} J.~L.,    {West} M.~J.,  2006, \apjl, 644,
  L21

\bibitem[\protect\citeauthoryear{{Ferrarese} \& {Merritt}}{{Ferrarese} \&
  {Merritt}}{2000}]{2000ApJ...539L...9F}
{Ferrarese} L.,  {Merritt} D.,  2000, \apjl, 539, L9

\bibitem[\protect\citeauthoryear{{Gavazzi}, {Treu}, {Rhodes}, {Koopmans},
  {Bolton}, {Burles}, {Massey} \& {Moustakas}}{{Gavazzi}
  et~al.}{2007}]{2007ApJ...667..176G}
{Gavazzi} R.,  {Treu} T.,  {Rhodes} J.~D.,  {Koopmans} L.~V.~E.,  {Bolton}
  A.~S.,  {Burles} S.,  {Massey} R.~J.,    {Moustakas} L.~A.,  2007, \apj, 667,
  176

\bibitem[\protect\citeauthoryear{{Gebhardt}, {Bender}, {Bower}, {Dressler},
  {Faber}, {Filippenko}, {Green}, {Grillmair}, {Ho}, {Kormendy}, {Lauer},
  {Magorrian}, {Pinkney}, {Richstone} \& {Tremaine}}{{Gebhardt}
  et~al.}{2000}]{2000ApJ...539L..13G}
{Gebhardt} K.,  {Bender} R.,  {Bower} G.,  {Dressler} A.,  {Faber} S.~M.,
  {Filippenko} A.~V.,  {Green} R.,  {Grillmair} C.,  {Ho} L.~C.,  {Kormendy}
  J.,  {Lauer} T.~R.,  {Magorrian} J.,  {Pinkney} J.,  {Richstone} D.,
  {Tremaine} S.,  2000, \apjl, 539, L13

\bibitem[\protect\citeauthoryear{{Graham}}{{Graham}}{2007}]{2007MNRAS.379..711G}
{Graham} A.~W.,  2007, \mnras, 379, 711

\bibitem[\protect\citeauthoryear{{Graham} \& {Driver}}{{Graham} \&
  {Driver}}{2007}]{2007ApJ...655...77G}
{Graham} A.~W.,  {Driver} S.~P.,  2007, \apj, 655, 77

\bibitem[\protect\citeauthoryear{{Graham}, {Erwin}, {Caon} \&
  {Trujillo}}{{Graham} et~al.}{2001}]{2001ApJ...563L..11G}
{Graham} A.~W.,  {Erwin} P.,  {Caon} N.,    {Trujillo} I.,  2001, \apjl, 563,
  L11

\bibitem[\protect\citeauthoryear{{Graham}, {Onken}, {Athanassoula} \&
  {Combes}}{{Graham} et~al.}{2011}]{2011MNRAS.412.2211G}
{Graham} A.~W.,  {Onken} C.~A.,  {Athanassoula} E.,    {Combes} F.,  2011,
  \mnras, 412, 2211

\bibitem[\protect\citeauthoryear{{Grogin} \& {Narayan}}{{Grogin} \&
  {Narayan}}{1996a}]{1996ApJ...464...92G}
{Grogin} N.~A.,  {Narayan} R.,  1996a, \apj, 464, 92

\bibitem[\protect\citeauthoryear{{Grogin} \& {Narayan}}{{Grogin} \&
  {Narayan}}{1996b}]{1996ApJ...473..570G}
{Grogin} N.~A.,  {Narayan} R.,  1996b, \apj, 473, 570

\bibitem[\protect\citeauthoryear{{G{\"u}ltekin}, {Richstone}, {Gebhardt},
  {Lauer}, {Tremaine}, {Aller}, {Bender}, {Dressler}, {Faber}, {Filippenko},
  {Green}, {Ho}, {Kormendy}, {Magorrian}, {Pinkney} \& {Siopis}}{{G{\"u}ltekin}
  et~al.}{2009}]{2009ApJ...698..198G}
{G{\"u}ltekin} K.,  {Richstone} D.~O.,  {Gebhardt} K.,  {Lauer} T.~R.,
  {Tremaine} S.,  {Aller} M.~C.,  {Bender} R.,  {Dressler} A.,  {Faber} S.~M.,
  {Filippenko} A.~V.,  {Green} R.,  {Ho} L.~C.,  {Kormendy} J.,  {Magorrian}
  J.,  {Pinkney} J.,    {Siopis} C.,  2009, \apj, 698, 198

\bibitem[\protect\citeauthoryear{{Haiman}, {Ciotti} \& {Ostriker}}{{Haiman}
  et~al.}{2004}]{2004ApJ...606..763H}
{Haiman} Z.,  {Ciotti} L.,    {Ostriker} J.~P.,  2004, \apj, 606, 763

\bibitem[\protect\citeauthoryear{{H{\"a}ring} \& {Rix}}{{H{\"a}ring} \&
  {Rix}}{2004}]{2004ApJ...604L..89H}
{H{\"a}ring} N.,  {Rix} H.-W.,  2004, \apjl, 604, L89

\bibitem[\protect\citeauthoryear{{Hu}}{{Hu}}{2008}]{2008MNRAS.386.2242H}
{Hu} J.,  2008, \mnras, 386, 2242

\bibitem[\protect\citeauthoryear{{Inada}, {Oguri}, {Falco}, {Broadhurst},
  {Ofek}, {Kochanek}, {Sharon} \& {Smith}}{{Inada}
  et~al.}{2008}]{2008PASJ...60L..27I}
{Inada} N.,  {Oguri} M.,  {Falco} E.~E.,  {Broadhurst} T.~J.,  {Ofek} E.~O.,
  {Kochanek} C.~S.,  {Sharon} K.,    {Smith} G.~P.,  2008, \pasj, 60, L27+

\bibitem[\protect\citeauthoryear{{Johansson}, {Burkert} \& {Naab}}{{Johansson}
  et~al.}{2009}]{2009ApJ...707L.184J}
{Johansson} P.~H.,  {Burkert} A.,    {Naab} T.,  2009, \apjl, 707, L184

\bibitem[\protect\citeauthoryear{{Kassiola} \& {Kovner}}{{Kassiola} \&
  {Kovner}}{1993}]{1993ApJ...417..450K}
{Kassiola} A.,  {Kovner} I.,  1993, \apj, 417, 450

\bibitem[\protect\citeauthoryear{{Kauffmann} \& {Haehnelt}}{{Kauffmann} \&
  {Haehnelt}}{2000}]{2000MNRAS.311..576K}
{Kauffmann} G.,  {Haehnelt} M.,  2000, \mnras, 311, 576

\bibitem[\protect\citeauthoryear{{Keeton}}{{Keeton}}{2003}]{2003ApJ...582...17K}
{Keeton} C.~R.,  2003, \apj, 582, 17

\bibitem[\protect\citeauthoryear{{Keeton} \& {Kochanek}}{{Keeton} \&
  {Kochanek}}{1998}]{1998ApJ...495..157K}
{Keeton} C.~R.,  {Kochanek} C.~S.,  1998, \apj, 495, 157

\bibitem[\protect\citeauthoryear{{Kochanek}}{{Kochanek}}{1995}]{1995ApJ...445..559K}
{Kochanek} C.~S.,  1995, \apj, 445, 559

\bibitem[\protect\citeauthoryear{{Kochanek}}{{Kochanek}}{1996}]{1996ApJ...466..638K}
{Kochanek} C.~S.,  1996, \apj, 466, 638

\bibitem[\protect\citeauthoryear{{Koopmans}, {Treu}, {Bolton}, {Burles} \&
  {Moustakas}}{{Koopmans} et~al.}{2006}]{2006ApJ...649..599K}
{Koopmans} L.~V.~E.,  {Treu} T.,  {Bolton} A.~S.,  {Burles} S.,    {Moustakas}
  L.~A.,  2006, \apj, 649, 599

\bibitem[\protect\citeauthoryear{{Kormann}, {Schneider} \&
  {Bartelmann}}{{Kormann} et~al.}{1994}]{1994A&A...284..285K}
{Kormann} R.,  {Schneider} P.,    {Bartelmann} M.,  1994, \aap, 284, 285

\bibitem[\protect\citeauthoryear{{Laor}}{{Laor}}{2001}]{2001ApJ...553..677L}
{Laor} A.,  2001, \apj, 553, 677

\bibitem[\protect\citeauthoryear{{Magorrian}, {Tremaine}, {Richstone},
  {Bender}, {Bower}, {Dressler}, {Faber}, {Gebhardt}, {Green}, {Grillmair},
  {Kormendy} \& {Lauer}}{{Magorrian} et~al.}{1998}]{1998AJ....115.2285M}
{Magorrian} J.,  {Tremaine} S.,  {Richstone} D.,  {Bender} R.,  {Bower} G.,
  {Dressler} A.,  {Faber} S.~M.,  {Gebhardt} K.,  {Green} R.,  {Grillmair} C.,
  {Kormendy} J.,    {Lauer} T.,  1998, \aj, 115, 2285

\bibitem[\protect\citeauthoryear{{Mao} \& {Witt}}{{Mao} \& {Witt}}{2011}]{mw11}
{Mao} S.,  {Witt} H.~J.,  2011, \mnras, submitted

\bibitem[\protect\citeauthoryear{{Mao}, {Witt} \& {Koopmans}}{{Mao}
  et~al.}{2001}]{2001MNRAS.323..301M}
{Mao} S.,  {Witt} H.~J.,    {Koopmans} L.~V.~E.,  2001, \mnras, 323, 301

\bibitem[\protect\citeauthoryear{{Maoz} \& {Rix}}{{Maoz} \&
  {Rix}}{1993}]{1993ApJ...416..425M}
{Maoz} D.,  {Rix} H.,  1993, \apj, 416, 425

\bibitem[\protect\citeauthoryear{{Marconi} \& {Hunt}}{{Marconi} \&
  {Hunt}}{2003}]{2003ApJ...589L..21M}
{Marconi} A.,  {Hunt} L.~K.,  2003, \apjl, 589, L21

\bibitem[\protect\citeauthoryear{{McKean}, {Browne}, {Jackson}, {Koopmans},
  {Norbury}, {Treu}, {York}, {Biggs}, {Blandford}, {de Bruyn}, {Fassnacht},
  {Mao}, {Myers}, {Pearson}, {Phillips}, {Readhead}, {Rusin} \&
  {Wilkinson}}{{McKean} et~al.}{2005}]{2005MNRAS.356.1009M}
{McKean} J.~P.,  {Browne} I.~W.~A.,  {Jackson} N.~J.,  {Koopmans} L.~V.~E.,
  {Norbury} M.~A.,  {Treu} T.,  {York} T.~D.,  {Biggs} A.~D.,  {Blandford}
  R.~D.,  {de Bruyn} A.~G.,  {Fassnacht} C.~D.,  {Mao} S.,  {Myers} S.~T.,
  {Pearson} T.~J.,  {Phillips} P.~M.,  {Readhead} A.~C.~S.,  {Rusin} D.,
  {Wilkinson} P.~N.,  2005, \mnras, 356, 1009

\bibitem[\protect\citeauthoryear{{McLure} \& {Dunlop}}{{McLure} \&
  {Dunlop}}{2001}]{2001MNRAS.327..199M}
{McLure} R.~J.,  {Dunlop} J.~S.,  2001, \mnras, 327, 199

\bibitem[\protect\citeauthoryear{{McLure} \& {Dunlop}}{{McLure} \&
  {Dunlop}}{2002}]{2002MNRAS.331..795M}
{McLure} R.~J.,  {Dunlop} J.~S.,  2002, \mnras, 331, 795

\bibitem[\protect\citeauthoryear{{Merritt} \& {Milosavljevi{\'c}}}{{Merritt} \&
  {Milosavljevi{\'c}}}{2005}]{2005LRR.....8....8M}
{Merritt} D.,  {Milosavljevi{\'c}} M.,  2005, Living Reviews in Relativity, 8,
  8

\bibitem[\protect\citeauthoryear{{Monaco}, {Salucci} \& {Danese}}{{Monaco}
  et~al.}{2000}]{2000MNRAS.311..279M}
{Monaco} P.,  {Salucci} P.,    {Danese} L.,  2000, \mnras, 311, 279

\bibitem[\protect\citeauthoryear{{More}, {McKean}, {Muxlow}, {Porcas},
  {Fassnacht} \& {Koopmans}}{{More} et~al.}{2008}]{2008MNRAS.384.1701M}
{More} A.,  {McKean} J.~P.,  {Muxlow} T.~W.~B.,  {Porcas} R.~W.,  {Fassnacht}
  C.~D.,    {Koopmans} L.~V.~E.,  2008, \mnras, 384, 1701

\bibitem[\protect\citeauthoryear{{Navarro}, {Frenk} \& {White}}{{Navarro}
  et~al.}{1997}]{1997ApJ...490..493N}
{Navarro} J.~F.,  {Frenk} C.~S.,    {White} S.~D.~M.,  1997, \apj, 490, 493

\bibitem[\protect\citeauthoryear{{Nipoti}, {Londrillo} \& {Ciotti}}{{Nipoti}
  et~al.}{2003}]{2003MNRAS.342..501N}
{Nipoti} C.,  {Londrillo} P.,    {Ciotti} L.,  2003, \mnras, 342, 501

\bibitem[\protect\citeauthoryear{{Novak}, {Faber} \& {Dekel}}{{Novak}
  et~al.}{2006}]{2006ApJ...637...96N}
{Novak} G.~S.,  {Faber} S.~M.,    {Dekel} A.,  2006, \apj, 637, 96

\bibitem[\protect\citeauthoryear{{Parker}, {Hoekstra}, {Hudson}, {van Waerbeke}
  \& {Mellier}}{{Parker} et~al.}{2007}]{2007ApJ...669...21P}
{Parker} L.~C.,  {Hoekstra} H.,  {Hudson} M.~J.,  {van Waerbeke} L.,
  {Mellier} Y.,  2007, \apj, 669, 21

\bibitem[\protect\citeauthoryear{{Peng}}{{Peng}}{2007}]{Peng2007}
{Peng} C.~Y.,  2007, \apj, 671, 1098

\bibitem[\protect\citeauthoryear{{Rix}, {de Zeeuw}, {Cretton}, {van der Marel}
  \& {Carollo}}{{Rix} et~al.}{1997}]{1997ApJ...488..702R}
{Rix} H.,  {de Zeeuw} P.~T.,  {Cretton} N.,  {van der Marel} R.~P.,
  {Carollo} C.~M.,  1997, \apj, 488, 702

\bibitem[\protect\citeauthoryear{{Robertson}, {Hernquist}, {Cox}, {Di Matteo},
  {Hopkins}, {Martini} \& {Springel}}{{Robertson}
  et~al.}{2006}]{2006ApJ...641...90R}
{Robertson} B.,  {Hernquist} L.,  {Cox} T.~J.,  {Di Matteo} T.,  {Hopkins}
  P.~F.,  {Martini} P.,    {Springel} V.,  2006, \apj, 641, 90

\bibitem[\protect\citeauthoryear{{Rusin}, {Keeton} \& {Winn}}{{Rusin}
  et~al.}{2005}]{2005ApJ...627L..93R}
{Rusin} D.,  {Keeton} C.~R.,    {Winn} J.~N.,  2005, \apjl, 627, L93

\bibitem[\protect\citeauthoryear{{Rusin} \& {Kochanek}}{{Rusin} \&
  {Kochanek}}{2005}]{2005ApJ...623..666R}
{Rusin} D.,  {Kochanek} C.~S.,  2005, \apj, 623, 666

\bibitem[\protect\citeauthoryear{{Rusin}, {Kochanek} \& {Keeton}}{{Rusin}
  et~al.}{2003}]{2003ApJ...595...29R}
{Rusin} D.,  {Kochanek} C.~S.,    {Keeton} C.~R.,  2003, \apj, 595, 29

\bibitem[\protect\citeauthoryear{{Rusin} \& {Ma}}{{Rusin} \&
  {Ma}}{2001}]{2001ApJ...549L..33R}
{Rusin} D.,  {Ma} C.-P.,  2001, \apjl, 549, L33

\bibitem[\protect\citeauthoryear{{Rusin} \& {Tegmark}}{{Rusin} \&
  {Tegmark}}{2001}]{RT01}
{Rusin} D.,  {Tegmark} M.,  2001, \apj, 553, 709

\bibitem[\protect\citeauthoryear{{Soker}}{{Soker}}{2009}]{2009MNRAS.398L..41S}
{Soker} N.,  2009, \mnras, 398, L41

\bibitem[\protect\citeauthoryear{{Tremaine}, {Gebhardt}, {Bender}, {Bower},
  {Dressler}, {Faber}, {Filippenko}, {Green}, {Grillmair}, {Ho}, {Kormendy},
  {Lauer}, {Magorrian}, {Pinkney} \& {Richstone}}{{Tremaine}
  et~al.}{2002}]{2002ApJ...574..740T}
{Tremaine} S.,  {Gebhardt} K.,  {Bender} R.,  {Bower} G.,  {Dressler} A.,
  {Faber} S.~M.,  {Filippenko} A.~V.,  {Green} R.,  {Grillmair} C.,  {Ho}
  L.~C.,  {Kormendy} J.,  {Lauer} T.~R.,  {Magorrian} J.,  {Pinkney} J.,
  {Richstone} D.,  2002, \apj, 574, 740

\bibitem[\protect\citeauthoryear{{Treu} \& {Koopmans}}{{Treu} \&
  {Koopmans}}{2002}]{2002ApJ...575...87T}
{Treu} T.,  {Koopmans} L.~V.~E.,  2002, \apj, 575, 87

\bibitem[\protect\citeauthoryear{{Trujillo}, {Erwin}, {Asensio Ramos} \&
  {Graham}}{{Trujillo} et~al.}{2004}]{2004AJ....127.1917T}
{Trujillo} I.,  {Erwin} P.,  {Asensio Ramos} A.,    {Graham} A.~W.,  2004, \aj,
  127, 1917

\bibitem[\protect\citeauthoryear{{Tu}, {Gavazzi}, {Limousin}, {Cabanac},
  {Marshall}, {Fort}, {Treu}, {P{\'e}llo}, {Jullo}, {Kneib} \& {Sygnet}}{{Tu}
  et~al.}{2009}]{2009A&A...501..475T}
{Tu} H.,  {Gavazzi} R.,  {Limousin} M.,  {Cabanac} R.,  {Marshall} P.~J.,
  {Fort} B.,  {Treu} T.,  {P{\'e}llo} R.,  {Jullo} E.,  {Kneib} J.-P.,
  {Sygnet} J.-F.,  2009, \aap, 501, 475

\bibitem[\protect\citeauthoryear{{Volonteri}, {Haardt} \& {Madau}}{{Volonteri}
  et~al.}{2003}]{2003ApJ...582..559V}
{Volonteri} M.,  {Haardt} F.,    {Madau} P.,  2003, \apj, 582, 559

\bibitem[\protect\citeauthoryear{{Winn}, {Rusin} \& {Kochanek}}{{Winn}
  et~al.}{2004}]{2004Natur.427..613W}
{Winn} J.~N.,  {Rusin} D.,    {Kochanek} C.~S.,  2004, \nat, 427, 613

\bibitem[\protect\citeauthoryear{{Wyithe} \& {Loeb}}{{Wyithe} \&
  {Loeb}}{2002}]{2002ApJ...581..886W}
{Wyithe} J.~S.~B.,  {Loeb} A.,  2002, \apj, 581, 886

\bibitem[\protect\citeauthoryear{{Yu}}{{Yu}}{2002}]{2002MNRAS.331..935Y}
{Yu} Q.,  2002, \mnras, 331, 935

\bibitem[\protect\citeauthoryear{{Yu} \& {Lu}}{{Yu} \&
  {Lu}}{2008}]{2008ApJ...689..732Y}
{Yu} Q.,  {Lu} Y.,  2008, \apj, 689, 732

\bibitem[\protect\citeauthoryear{{Yu} \& {Tremaine}}{{Yu} \&
  {Tremaine}}{2002}]{2002MNRAS.335..965Y}
{Yu} Q.,  {Tremaine} S.,  2002, \mnras, 335, 965

\bibitem[\protect\citeauthoryear{{Zhang}, {Jackson}, {Porcas} \&
  {Browne}}{{Zhang} et~al.}{2007}]{2007MNRAS.377.1623Z}
{Zhang} M.,  {Jackson} N.,  {Porcas} R.~W.,    {Browne} I.~W.~A.,  2007,
  \mnras, 377, 1623

\end{thebibliography}

\end{document}